\documentclass[preprintnumbers,amsmath,amssymb,twocolumn,aps,prl,10pt]{revtex4-1}

\usepackage{graphicx,comment}% Include figure files
\usepackage{epstopdf}
\usepackage{dcolumn}% Align table columns on decimal point
\usepackage{bm}% bold math
\usepackage[T1]{fontenc}
\usepackage[latin1]{inputenc}
\usepackage{enumerate}
\usepackage{color}
\usepackage{verbatim}
\usepackage{hyperref}
\usepackage[all]{hypcap}
\usepackage{mathtools}\mathtoolsset{centercolon}
\usepackage{titlesec}
\usepackage[export]{adjustbox}

\titleformat{\section}[runin]
  {\normalfont\bfseries}
  {\thesection}
  {1em}
  {\addperiod}
\newcommand{\addperiod}[1]{#1.}
\titlespacing*{\section}{0pt}{1ex}{1ex}

\DeclareMathOperator{\Tr}{Tr}
\DeclareMathOperator{\spn}{span}
\DeclareMathOperator{\sel}{select}
\newcommand{\sV}{\sel(V)}

\newcommand{\intv}{M}
\newcommand{\ket}[1]{\left| #1\right\rangle}        % ket vector
\newcommand{\bra}[1]{\left\langle #1\right|}        % bra vector

\newcommand{\eq}[1]{\hyperref[eq:#1]{(\ref*{eq:#1})}}

\newcommand{\app}[1]{\hyperref[app:#1]{Appendix~\ref*{app:#1}}}
\begin{document}

\title{Simulating Hamiltonian dynamics with a truncated Taylor series}

\author{Dominic W. Berry${}^1$, Andrew M. Childs${}^{2,3,4,5}$, Richard Cleve${}^{2,5,6}$, Robin Kothari${}^{2,6,7}$, and Rolando D. Somma${}^8$}
\affiliation{${}^1$Department of Physics and Astronomy, Macquarie University, Sydney, NSW 2109, Australia
\\
${}^2$Institute for Quantum Computing, University of Waterloo, ON N2L 3G1, Canada 
\\
${}^3$Department of Combinatorics \& Optimization, University of Waterloo, ON N2L 3G1, Canada 
\\
${}^4$Department of Computer Science, Institute for Advanced Computer Studies, and Joint Center for Quantum Information and Computer Science, University of Maryland, College Park, MD 20910, USA 
\\
${}^5$Canadian Institute for Advanced Research, Toronto, Ontario M5G 1Z8, Canada
\\
${}^6$School of Computer Science, University of Waterloo, ON N2L 3G1, Canada 
\\
${}^7$Center for Theoretical Physics, Massachusetts Institute of Technology, Cambridge, MA 02139, USA 
\\
${}^8$Theoretical Division, Los Alamos National Laboratory, Los Alamos, NM 87545, USA
}
\date{\today}

\begin{abstract}
We describe a simple, efficient method for simulating Hamiltonian dynamics on a quantum computer by approximating the truncated Taylor series of the evolution operator.  Our method can simulate the time evolution of a wide variety of physical systems.
As in another recent algorithm, the cost of our method depends only logarithmically on the inverse of the desired precision, which is optimal.
However, we simplify the algorithm and its analysis by
using a method for implementing linear combinations of unitary operations to directly apply the truncated Taylor series.
\end{abstract}

\preprint{LA-UR-14-22745, MIT-CTP \#4618}

\maketitle

%%%%%%%%%%%%%%%%%%%%%%%%%%%%%%%%%%%%%%%%%%%%%%%
%%%%%%%%%%%%%%%%%%%%%%%%%%%%%%%%%%%%%%%%%%%%%%%
% \section{Introduction}

One of the main motivations for quantum computers is their ability to efficiently simulate the dynamics of quantum systems \cite{Fey82}.
Since the mid-1990s, many algorithms have been developed to simulate Hamiltonian dynamics on a quantum computer \cite{Llo96,AT03,Chi04,BAC07,CK11,WBH11,WBH10,Ch09,BC12,PQS11,CW12}, with applications to problems such as simulating spin models \cite{SOGKL02} and quantum chemistry \cite{WBCHT14,BMWAW14,BLA14,MBLA14}.
While it is now well known that quantum computers can efficiently simulate Hamiltonian dynamics, ongoing work has improved the performance and expanded the scope of such simulations.

Recently, we introduced a new approach to Hamiltonian simulation with exponentially improved performance as a function of the desired precision \cite{BCC14}.
Specifically, we presented a method to simulate a $d$-sparse, $n$-qubit Hamiltonian $H$ acting for time $t>0$, within precision $\epsilon>0$, using $O(\tau \log({\tau}/{\epsilon}) / \log\log({\tau}/{\epsilon}))$ queries to $H$ and $O(n \tau \log^2({\tau}/{\epsilon}) / \log\log({\tau}/{\epsilon}))$ additional two-qubit gates,
where $\tau := d^2 \| H \|_{\max} t$.
This dependence on $\epsilon$ is exponentially better than all previous approaches to Hamiltonian simulation, and the number of queries to $H$ is optimal \cite{BCC14}.
(For simplicity, we refer to combinations of logarithms like those in the above expressions as logarithmic.)
Roughly speaking, doubling the number of digits of accuracy only doubles the complexity.

The simulation algorithm of \cite{BCC14} is 
indirect, appealing to an unconventional model of query complexity.
In this paper, we describe and analyze a simplified approach to Hamiltonian simulation with the same cost as the method of \cite{BCC14}.
The new approach is easier to understand, and the reason for the logarithmic cost dependence on $\epsilon$ is immediate.
The new approach decomposes the Hamiltonian into a linear combination of unitary operations.  Unlike the algorithm of \cite{BCC14}, these terms need not be self-inverse, so the algorithm is efficient for a larger class of Hamiltonians.  The new approach is also simpler to analyze: we give a self-contained presentation in four pages.

The main idea of the new approach is to implement the truncated Taylor series of the evolution operator.  Similarly to previous approaches for implementing linear combinations of unitary operators \cite{CW12,SOGKL02}, the various terms of the Taylor series can be implemented by introducing an ancillary superposition and performing controlled operations.
The time evolution is broken up into \emph{segments}, each of which is short enough that the evolution can be accurately approximated using a number of terms in the Taylor series that is logarithmic in $1/\epsilon$.
Each segment is then performed using \emph{oblivious amplitude amplification} \cite{BCC14}.
To make this work in the present context, we show that a more powerful \emph{robust} version of amplitude amplification holds, where the target operation need not be unitary (and our proof of this is simpler than the proof for a weaker version in \cite{BCC14}).
The complexity of the method is essentially given by the order at which the series is truncated times the number of segments.

Our algorithm can be applied to simulate various models of realistic systems, including 
systems of spins or fermions.
It can also be used to implement other quantum algorithms \cite{AT03, CCJY09, HHL09}.
We focus on the case of   time-independent Hamiltonians for simplicity, but we also outline a straightforward application of our results to time-dependent Hamiltonians. 

Specifically, we present a quantum algorithm that simulates the time evolution of a finite-dimensional Hamiltonian of the form
\begin{align}\label{eq:original-sum}
H = \sum_{\ell=1}^L \alpha_\ell H_\ell ,
\end{align}
where each $H_\ell$ is unitary and a mechanism is available for implementing that unitary.  Any Hamiltonian can be decomposed as a linear combination of unitary matrices, and many have decompositions into a small number of terms that are easy to implement.  For example, local Hamiltonians can be decomposed into a sum of tensor products of Pauli matrices where each term acts nontrivially on a constant number of qubits.  More generally, we can treat any Hamiltonian that is a sum of polynomially many tensor products of Pauli operators, even if the terms have high weight (as occurs after a Jordan-Wigner transformation of fermionic operators into Pauli matrices).  
The method can also be applied more broadly, e.g., to sparse Hamiltonians.
The overall simulation complexity is $T := (\alpha_1+\cdots+\alpha_L)t$ times a factor that is logarithmic in $1/\epsilon$ and the other parameters of the simulation.

%%%%%%%%%%%%%%%%%%%%%%%%%%%%%%%%%%%%%%%%%%%%%%%
%%%%%%%%%%%%%%%%%%%%%%%%%%%%%%%%%%%%%%%%%%%%%%%
\section{Summary of method}

Suppose we wish to simulate the evolution under a Hamiltonian $H$ for time $t$, 
\begin{equation}
U := \exp(-iHt) ,
\end{equation}
within error $\epsilon$.
We divide the evolution time into $r$ segments of length $t/r$.
Within each segment, the evolution can be approximated as
\begin{equation}
\label{eq:truncated}
U_r := \exp(-iHt/r) \approx \sum_{k=0}^{K} \frac{1}{k!} (-iHt/r)^k ,
\end{equation}
where the Taylor series is truncated at order $K$.
With $r$ segments, the accuracy required for each segment is $\epsilon/r$.
To obtain this accuracy (provided $r \ge \|H\|t$), we can choose
\begin{equation}
K = O\biggl( \frac{\log(r/\epsilon)}{\log\log(r/\epsilon)} \biggr).
\end{equation}
The overall complexity is essentially given by the number of segments $r$ times $K$.

The oblivious amplitude amplification procedure of \cite{BCC14} enables us to deterministically implement a sum of unitary operators, provided that sum is unitary.
Here we wish to implement the sum of operators corresponding to the truncated Taylor series \eq{truncated}.
The powers of $H$ are not themselves unitary, but we can expand the series using the form \eq{original-sum} of the Hamiltonian.
Then the truncated sum in \eq{truncated} can be expanded as
\begin{align}
U_r \approx \sum_{k=0}^{K}
\sum_{\ell_1,\ldots,\ell_k=1}^L
\frac{(-it/r)^k}{k!}\alpha_{\ell_1} \cdots \alpha_{\ell_k} \, H_{\ell_1} \cdots H_{\ell_k} \label{eq:explict} ,
\end{align}
where, without loss of generality, we can set each $\alpha_\ell > 0$.
This expression has a form that has been investigated extensively in~\cite{Kothari14}, namely
\begin{align}\label{eq:abstract-finite}
\widetilde U &= \sum_{j=0}^{m-1} \beta_j V_j ,
\end{align}
where $\beta_j > 0$ and where each $V_j$ is a unitary that corresponds to some
$(-i)^k H_{\ell_1} \cdots H_{\ell_k}$.

Next we explain how to implement a sum of unitary operators such as \eq{abstract-finite}.
This procedure would work exactly if the sum were unitary.
While the sum is not exactly unitary, it is close to unitary, and the error can be bounded.

We begin by assuming that there is a mechanism available for implementing each unitary $V_j$ (and we address the details of this mechanism later).
We abstract this mechanism as a unitary operation $\sV$ such that, 
for any $j \in \{0,1,\dots,m-1\}$ and any state $\ket{\psi}$, 
\begin{align}
\sV\ket{j}\ket{\psi} = \ket{j}V_j\ket{\psi}.
\end{align}

To simulate $\widetilde U$, we first implement an $m$-dimensional unitary $B$ to prepare the ancillary state 
\begin{align}
\label{eq:Aaction}
B\ket{0} = \frac{1}{\sqrt s}\sum_{j=0}^{m-1} \sqrt{\beta_j}\ket{j},
\end{align}
where we define $s:= \sum_{j=0}^{m-1} \beta_j$.
If we define 
\begin{align}
W := (B^{\dag} \otimes \openone)(\sV)(B \otimes \openone),
\end{align}
then 
\begin{align}\label{eq:w}
W\ket{0}\ket{\psi} &= \frac{1}{s}\ket{0} \widetilde U\ket{\psi} + \sqrt{1-\frac{1}{s^2}}\ket{\Phi}
\end{align}
for some $\ket{\Phi}$ whose ancillary state is supported in the subspace orthogonal to $\ket 0$.
In other words, applying
the projector $P:=\ket{0}\! \bra{0}\otimes \openone$, 
\begin{align}
P W \ket{0}\ket{\psi} =
\frac{1}{s}\ket{0}\widetilde U\ket{\psi}.
\label{eq:Waction}
\end{align}

The value of $s$ can be adjusted by choosing the size of the segments.
In particular, we aim for $s=2$, which fits the framework of oblivious amplitude amplification introduced in \cite{BCC14}.
There it is shown that if $\widetilde U$ is unitary then it can be exactly implemented by interleaving $W$ and $W^{\dag}$ with the $m$-dimensional (ancilla) reflection $R := \openone - 2P$.  In particular, if $\widetilde U$ is unitary
and $A := -W R W^{\dag} R W$,
then 
\begin{align}\label{eq:oaa}
A\ket{0}\ket{\psi} &= \ket{0}\widetilde U\ket{\psi}.
\end{align}

%%%%%%%%%%%%%%%%%%%%%%%%%%%%%%%%%%%%%%%%%%%%%%%
%%%%%%%%%%%%%%%%%%%%%%%%%%%%%%%%%%%%%%%%%%%%%%%
\section{Effect of nonunitarity}

The main approximation used in our construction is in
truncating the Taylor series at a suitably large order $K$, which gives a good (but nonunitary) approximation $\widetilde U$ of $U_r$. Due to this nonunitarity, we cannot directly use oblivious amplitude amplification as proved in \cite{BCC14}; instead, we prove a robust version of  oblivious amplitude amplification that works even when $\widetilde U$ is only close to a unitary matrix.

To prove this, first observe that
\begin{align}
\label{eq:OAA2}
PA \ket 0 \ket \psi = (3 P W -4 P W P W^\dagger P W ) \ket 0 \ket \psi ,
\end{align}
where we used 
unitarity of $W$,
$P^2=P$, and $P \ket 0 \ket \psi= \ket 0 \ket \psi$.
In our construction, $PWP = (\ket 0 \bra 0 \otimes \widetilde U)/s$ as can be seen from \eq{Waction}. Thus,  by \eq{OAA2}, we obtain (for general $\widetilde U$)
\begin{align}
\label{eq:Paction}
PA\ket{0}\ket{\psi} &= \ket 0 \biggl( \frac{3}{s} \widetilde U - \frac{4}{s^3} \widetilde U \widetilde U ^\dagger \widetilde U \biggr)   \ket \psi .
\end{align}
This generalizes a step of oblivious amplitude amplification [as in \eq{oaa}] to the case of nonunitary $\widetilde U$ and $s\ne 2$.
It also enables us to bound the error in oblivious amplitude amplification due to these factors.
Provided that $|s-2|=O(\delta)$ and $\| \widetilde U - U_r \| =O( \delta)$, i.e., that the conditions of oblivious amplitude amplification are true to order $\delta$, then $\| \widetilde U \widetilde U^\dagger - \openone \| = O(\delta)$ and \eq{Paction} implies
\begin{align}
\label{eq:OAAproj}
\| {P A\ket{0}\ket{\psi} -\ket{0}  U_r \ket{\psi}} \| = O(\delta),
\end{align}
which is an approximate analogue of \eq{oaa} up to order $\delta$.
In turn, this means that
\begin{align}
\label{eq:OAAfinalstate}
\| {\Tr_{\rm anc}(P A\ket{0}\ket{\psi}) -  U_r \ket{\psi}}\bra{\psi}U_r^\dag \| = O(\delta) ,
\end{align}
where $\Tr_{\rm anc}$ denotes the trace over the ancilla.  Thus if we initialize the ancilla for each segment and discard it afterward, the error is $O(\delta)$.
Then the overall error for all segments is $O(r\delta)$, so we can take $\delta=O(\epsilon/r)$.
In contrast, in \cite{BCC14} the error for each segment is $O(\sqrt\delta)$ (though this does not affect the overall complexity, which is logarithmic in $\delta$).

%%%%%%%%%%%%%%%%%%%%%%%%%%%%%%%%%%%%%%%%%%%%%%%
%%%%%%%%%%%%%%%%%%%%%%%%%%%%%%%%%%%%%%%%%%%%%%%

\section{Hamiltonian simulation algorithm}

For Hamiltonian simulation, the sum \eq{abstract-finite} is of the specific form given in \eq{explict}.
It is then convenient to define the index set
\begin{align}
\hspace*{-2.13mm}
J := \{(k,\ell_1,\dots,\ell_k) : k \in \mathbb{N}, \ell_1,\dots,\ell_k \in \{1,\dots,L\} \}
\end{align}
and, with respect to these indices, we let $\beta_{(k,\ell_1,\dots,\ell_k)} := 
[(t/r)^k / k!]\alpha_{\ell_1} \cdots \alpha_{\ell_k}$ and $V_{(k,\ell_1,\dots,\ell_k)} := (-i)^k H_{\ell_1} \cdots H_{\ell_k}$.
Then $U_r = \sum_{j \in J} \beta_j V_j $ with $\beta_j >0$.

Recall that $T = (\alpha_1+\cdots+\alpha_L)t$.
Taking $r$ such that $T/r=\ln(2)$, we obtain $\sum_{j \in J} \beta_j = 2$.
If $T$ is not a multiple of $\ln(2)$, then we can take $r=\lceil T/\ln(2) \rceil$, in which case the final segment has $s<2$.
We can compensate for this using an ancilla qubit as in \cite{BCC14}.

Next we set
\begin{equation}\label{eq:bound-on-k}
K = O\biggl(\frac{\log(T/\epsilon)}{\log\log(T/\epsilon)}\biggr)
\end{equation}
so that 
\begin{equation}
\sum_{k=K+1}^{\infty}\frac{\ln(2)^k}{k!} \le \epsilon /r,
\end{equation}
and define the truncated index set
\begin{align}
\widetilde{J} := \{(k,\ell_1,\dots,\ell_k) \in J : k \le K\}.
\end{align}
We obtain $\widetilde U = \sum_{j \in \tilde J} \beta_j V_j$ and,
according to \eq{Aaction}, $B$ is a unitary that acts on 
$\spn\{\ket{j} : j \in \widetilde{J}\}$ as
\begin{align}
{B}\ket{0} = \frac{1}{\sqrt{ {s}}}\sum_{j \in \widetilde{J}} \sqrt{\beta_j}\ket{j}.
\end{align}
The normalization constant is $s=\sum_{j \in \tilde J} \beta_j = \sum_{k=0}^{K} \frac{1}{k!} \ln(2)^k$.
Our choice of $K$ implies $|s-2| \le \epsilon/r$.

When the length of the simulation is a multiple of $\ln(2)$,
the Hamiltonian simulation algorithm is simply a sequence of $r$ steps,
each associated with a segment. At each step, the $m$-dimensional ancilla is initialized in $\ket 0$,
then $A$ is implemented, and the ancilla is discarded. 
The initial state is $\ket 0 \ket \psi$. 
To show that the overall error is $\le \epsilon$, it suffices to show
that the final state, after tracing out the ancillas, is $\epsilon$-close to $U \ket \psi$, where
$U$ is the evolution operator induced by $H$ for time $t$.
Our choice of $K$ implies that $\delta=O(\epsilon/r)$ in \eq{OAAfinalstate}.
Thus the error in the state after tracing out the ancillas is $O(\epsilon/r)$ for each segment.
Using subadditivity and appropriately selecting constants, the final error is $\le \epsilon$.

%%%%%%%%%%%%%%%%%%%%%%%%%%%%%%%%%%%%%%%%%%%%%%%
%%%%%%%%%%%%%%%%%%%%%%%%%%%%%%%%%%%%%%%%%%%%%%%

\section{Circuit constructions and gate counts}

The basic step of our method is the unitary $A$.
By definition, the cost of $A$ is the cost of two instances of $\sV$, one of $\sV^{\dag}$, and three each of $B$ and of $B^{\dag}$.
The complexity of implementing $\sV$ is $K$ times the complexity of implementing any of the individual $H_\ell$.

First we consider the number of ancillary qubits that are required to store the basis states $\ket{k}\ket{\ell_1}\cdots\ket{\ell_k}$ where $0 \le k \le K$ and, for each $\ell_i$, $1 \le \ell_i \le L$.
For the first register, we use a unary encoding, representing $\ket k$ by the state $\ket{1^{k}0^{K-k}}$, so the first register consists of 
$K = O(\log(T/\epsilon)/\log\log(T/\epsilon))$ qubits.
We store $K$ registers with the values $\ell_i$, but those after position $k$ are ignored when the first register is in state $\ket{k}\equiv \ket{1^{k}0^{K-k}}$.
Each of these registers contains $\log(L)$ qubits, so the total number of qubits required for the ancillary state is
\begin{align}
\label{eq:noqubits}
O\biggl(\frac{\log(L)\log(T/\epsilon)}{\log\log(T/\epsilon)}\biggr).
\end{align}

Now we consider the number of elementary (1- and 2-qubit) gates required to implement our construction.
The operation $B$ is a tensor product of unitary operations acting on each of the $K+1$ registers.
The first such unitary maps $\ket{0^K}$ to the normalized version of 
$\sum_{k=0}^K \sqrt{t^k/k!}\ket{1^k0^{K-k}}$.
This is easily achieved in $O(K)$ gates, using a rotation on the first qubit, followed by rotations on qubits $k=2$ to $K$
controlled by qubit $k-1$.
The subsequent unitaries map 
$\ket{0}$ to the normalized version of $\sum_{\ell=1}^L \sqrt{\alpha_{\ell}}\ket{\ell}$, which has a gate cost of $O(L)$~\cite{SBM06}.
Therefore the total gate cost of implementing $B$ is 
\begin{align}
O\biggl(L\frac{\log(T/\epsilon)}{\log\log(T/\epsilon)}\biggr).
\end{align}

The other basic component is the $\sV$ operation, which maps states of the form 
$\ket{k}\ket{\ell_1}\cdots\ket{\ell_k}\ket{\psi}$ to
$\ket{k}\ket{\ell_1}\cdots\ket{\ell_k}(-i)^k H_{\ell_1}\cdots H_{\ell_k}\ket{\psi}$.
This may be implemented using $K$ controlled-$\sel(H)$ operations that map states of the form $\ket{b}\ket{\ell}\ket{\psi}$ to 
$\ket{b}\ket{\ell}(-iH_{\ell})^{b}\ket{\psi}$.
The $\kappa$th controlled-$\sel(H)$ operation acts on the $\kappa$th qubit of $\ket k$ (which is encoded in unary) and the $\kappa$th register encoding $\ell_\kappa$.
This operation performs $-iH_{\ell_\kappa}$ on $\ket{\psi}$ provided $\kappa\le k$.

We next explain how to implement each controlled-$\sel(H)$ operation with $O(L(n + \log L))$ gates in the special case where each $H_{1},\dots,H_{L}$ is expressible as a tensor product of Pauli gates.
A controlled-$\sel(H)$ operation can be decomposed into a sequence of  
controlled-$H_{\ell}$ operations (with $\ell$ running from $1$ to $L$), each of which is controlled by two registers: the original control register (controlled by state $\ket{1}$) and the select register (controlled by state $\ket{\ell}$).
Each of these can in turn be implemented by combining an $O(\log L)$-qubit generalized Toffoli gate~\cite{BCD+95} [at cost $O(\log L)$] to set a control qubit and a controlled $n$-fold tensor product of Pauli gates [at cost $O(n)$].
It follows that the total cost of all controlled-$\sel(H)$ operations can be bounded by 
\begin{align}
\label{eq:gatecount-segment}
O\biggl(\frac{L(n + \log L)\log(T/\epsilon)}{\log\log(T/\epsilon)}\biggr)
\end{align}
for the $\sel(V)$ gate. The total number of gates for the entire simulation for time $t$
results from multiplying \eq{gatecount-segment} by $r=O(T)$, the number of segments. 

Alternatively, suppose the Hamiltonian is a sparse matrix given by an oracle.  Then we obtain the same overall scaling of the number of gates for a segment as in \cite{BCC14},
\begin{equation}
\label{eq:gatecount}
O\biggl( n \frac{\log^2(T/\epsilon)}{\log\log(T/\epsilon)} \biggr).
\end{equation}
In this scenario, the Hamiltonian is approximated by a sum of equal-size parts, with
$\alpha_\ell=\gamma$ for all $\ell$.  Then $\widetilde H = \gamma \sum_{\ell=1}^{L} V_\ell$ approximates the true Hamiltonian $H$ with error $O(\gamma)$.
To obtain overall error in the evolution $\le \epsilon$, $\gamma$ should be $\Theta(\epsilon/t)$.
The number of unitaries in the decomposition is $L=O(d^2 \| H \|_{\max}/\gamma)=O(d^2 \| H \|_{\max}t/\epsilon)$.
We have $T = t \sum_\ell \alpha_\ell = O(d^2\|H\|_{\max}t)$, so $L=O(T/\epsilon)$.
With all $\alpha_\ell$ equal, each ancilla register $\ket{\ell_i}$ can be prepared with complexity $\log L$.
For each segment, we multiply this by $K$ (for the number of these registers).
The 1-sparse self-inverse operations have complexity $O(n)$, but there is no need to perform an explicit sequence of $L$ controlled-$H_\ell$ operations when the Hamiltonian is given by an oracle.
This yields the gate complexity given in \eq{gatecount}.

%%%%%%%%%%%%%%%%%%%%%%%%%%%%%%%%%%%%%%%%%%%%%%%
%%%%%%%%%%%%%%%%%%%%%%%%%%%%%%%%%%%%%%%%%%%%%%%
\section{Time-dependent Hamiltonians}

A similar quantum algorithm can be applied to the time-dependent case if we consider the time-ordered exponential
\begin{equation}
U(t) := \mathcal T \exp \left\{ -i \int_0^t \mathrm{d}t' \, H(t') \right\} ,
\end{equation}
where $\mathcal T$ is the time-ordering operator.
As in the time-independent case, we break the evolution time into $r$ segments of size $t/r$ and,
with no loss of generality, we consider the evolution induced by the first segment $U_r := U(t/r)$.
This evolution can be approximated by
\begin{align}
\label{eq:EvolApprox1}
 \sum_{k=0}^{K} \frac{(-i)^k}{k!} \int_0^{t/r} \mathrm{d}\boldsymbol{t} \, \mathcal T  H(t_k) \ldots  H(t_{1}) .
\end{align}

Rather than directly implementing \eq{EvolApprox1}, 
we discretize the integral into $\intv$ steps and
approximate $U_r$ by
\begin{align}
\label{eq:EvolApprox2}
\widetilde U = \sum_{k=0}^{K} \frac{(-i t/r)^k}{M^k k!} \sum_{j_1,\ldots,j_k=0}^{\intv-1} \mathcal T H(t_{j_k}) \ldots H(t_{j_1}) .
\end{align}
The discrete times are $t_j=(j/M)t/r$.
Each of the individual Hamiltonians $H(t_{j_\ell})$ can then be expanded as a sum of $L$ unitaries,
so \eq{EvolApprox2}
is a sum of unitaries.
Thus $\widetilde U$ can be implemented similarly as in the time-independent case.
In this case, however, we need additional ancillas to encode the discrete times as $\ket{t_{j_1}} \cdots \ket{t_{j_k}}$.
Overall, for each segment we use one ancillary register to encode $k$, the order of the term; $K$ ancillary registers to encode $\ell_1,\ldots, \ell_k$;
and $K$ ancillas of size $\log(M)$ to encode the discrete times. 
The value of $M$ must be chosen sufficiently large that the overall error is at most $\epsilon$.
This value is polynomial in $1/\epsilon$ and $h' := \max_t \|\frac{\mathrm{d}}{\mathrm{d}t} H(t)\|$.
As the complexity is logarithmic in $M$, this only yields a factor logarithmic in $1/\epsilon$ and $h'$.

If $K$ is sufficiently large that $\widetilde U$ is almost unitary, we can use oblivious amplitude amplification to obtain a state within $O(\epsilon/r)$ of $U_r\ket{\psi}$.  By repeating this $r$ times (with the corresponding evolution times)
we can prepare the final state $U(t)\ket{\psi}$ to within error $\epsilon$.

\section{Conclusions}

We presented a simple quantum algorithm for Hamiltonian simulation that achieves the same complexity as in~\cite{BCC14}.
The new method is based on directly approximating the Taylor series for the evolution operator.
The logarithmic complexity in $\epsilon$ results because the $K$th-order approximation to the series has error of order $\frac{1}{K!}$.
Our method is generic: it does not treat the terms of the Hamiltonian in any particular order and does not require any special structural relationships among them.
Beyond Hamiltonian simulation, we expect that tools such as robust oblivious amplitude amplification and linear combinations of unitary operations will be more broadly useful in constructing quantum algorithms.

%%%%%%%%%%%%%%%%%%%%%%%%%%%%%%%%%%%%%%%%%%%%%%%
%%%%%%%%%%%%%%%%%%%%%%%%%%%%%%%%%%%%%%%%%%%%%%%

\section{Acknowledgments}

We thank Emmanuel Knill, John Preskill, and Nathan Wiebe for discussions.
D.W.B.\ was supported by ARC grant FT100100761.
R.S.\ acknowledges support from AFOSR through Grant No.\ FA9550-12-1-0057.
This work was also supported by ARO grant Contract Number W911NF-12-0486, Industry Canada, and NSERC.

%%%%%%%%%%%%%%%%%%%%%%%%%%%%%%%%%%%%%%%%%%%%%%%
%%%%%%%%%%%%%%%%%%%%%%%%%%%%%%%%%%%%%%%%%%%%%%%

\end{document}